 \documentclass[preprint2]{aastex}

\slugcomment{To appear in PASP}

\shorttitle{First Sources of Light}
\shortauthors{Bromm}

\begin{document}


\title{The First Sources of Light}


\author{Volker Bromm\altaffilmark{1} }
\affil{Harvard-Smithsonian Center for Astrophysics, 60 Garden Street,
    Cambridge, MA 02138;}
\email{vbromm@cfa.harvard.edu}


\altaffiltext{1}{2002 ASP Trumpler Award winner for an outstanding Ph.D. thesis,
carried out at Yale University.}


\begin{abstract}
I review recent progress in understanding the formation of the
first stars and quasars. The initial conditions for their emergence
are given by the now firmly established model of cosmological
structure formation. Numerical simulations of the collapse and
fragmentation of primordial gas indicate that the first stars
formed at redshifts $z\simeq 20 - 30$, and that they were predominantly
very massive, with $M_{\ast}\ga 100 M_{\odot}$. Important uncertainties,
however, remain. Paramount among them is the accretion process, which
builds up the final stellar mass by incorporating part of the
diffuse, dust-free envelope into the central protostellar core.
The first quasars, on the other hand, are predicted to have formed
later on, at $z\sim 10$, in more massive dark matter halos, with
total masses, $\sim 10^8 M_{\odot}$, characteristic of dwarf galaxies.
\end{abstract}



\section{INTRODUCTION}

How and when did the first sources of light form in the universe?
Within the framework of modern cosmology, we have learned that
the first stars and quasars formed at the end of the so-called
``dark ages'', a few $10^{8}$~yr after the big bang (e.g.,
Barkana \& Loeb 2001; Miralda-Escud\'{e} 2003).
The cosmic dark ages began $\sim$~500,000~yr after the big bang when 
the photons of the cosmic microwave background (CMB) were emitted.
At this time, the CMB photons redshifted into the infrared wavelength
band so that they
were no longer energetic enough to ionize hydrogen
atoms, and could henceforth propagate freely.
From the viewpoint of a human observer, the universe descended
into a state of complete darkness, that lasted until the
cosmic Renaissance of first light occurred much later on.

At these early times, the universe did not yet contain any gravitationally
bound structures, and its state was simple enough to allow a description
in terms of precise, linear physics (e.g., Hu \& Dodelson 2002).
We know from observations of the highest redshift quasars and
galaxies that the dark ages must have ended prior to
$z\sim 7$ (e.g., Hu et al. 2002; Fan et al. 2003). Whereas we can 
directly probe the state of the universe at $z\sim 1000$, by
measuring CMB anisotropies, and at $z\la 6$, by observing
quasar absorption lines and Lyman-$\alpha$ emission from galaxies, we
know little about the formative epoch in between. Thus, this era
is the final frontier of observational cosmology, and
the crucial question is (e.g., Barkana \& Loeb 2001; Loeb \& Barkana 2001;
Miralda-Escud\'{e}
2003): {\it How and when did the dark ages end?}

Within the now firmly established model of cosmological structure
formation (e.g., Spergel et al. 2003), the first stars are predicted
to have formed at redshifts $z\simeq 20 - 30$ in dark matter (DM) halos
of total mass $\sim 10^{6}M_{\odot}$ (e.g., Couchman \& Rees 1986;
Ostriker \& Gnedin 1996; Tegmark et al. 1997; Abel et al. 1998;
Yoshida et al. 2003a). 
Numerical simulations of the collapse and fragmentation of metal-free
gas, assumed to contain no dynamically significant magnetic fields yet,
have indicated that the first stars, the so-called Population~III
(e.g., Bond 1981), were predominantly very massive, with $M_{\ast}
\ga 100 M_{\odot}$ (e.g., Bromm, Coppi, \& Larson 1999, 2002;
Abel, Bryan, \& Norman 2000, 2002; Nakamura \& Umemura 2001).
Many uncertainties, however, remain. Two key problems are:
(i) The physics of accretion from a dust-free envelope, 
which determines the
final stellar mass, is not yet well understood (e.g., Omukai \& 
Palla 2001, 2003; Tan \& McKee 2003).
(ii) A complete theory of primordial star formation would allow us
to predict the exact functional form of the Population~III
initial mass function (IMF). Again, a reliable determination
is still eluding us (see, e.g., Nakamura \& Umemura 2001, 2002;
Omukai \& Yoshii 2003 for first attempts). I will address both
issues in \S 3.

The first quasars, on the other hand, powered by accretion onto
supermassive black holes (SMBHs), are expected to have formed later
on, at $z\sim 10$, in more massive, dwarf-sized systems (e.g., Haiman \&
Loeb 2001; Bromm \& Loeb 2003a). Numerical simulations of high-redshift
SMBH formation indicate that these could have formed only {\it after}
a significant episode of previous star formation 
(e.g., Bromm \& Loeb 2003a; Di Matteo et al. 2003).
The long-standing question (e.g., Silk \& Rees 1998) of what came
first, stars or quasars, would then be answered in favor of the former.

The formation of the first luminous sources had important implications
for the evolution of the intergalactic medium (IGM).
Massive Population~III stars were efficient producers of
ionizing photons (e.g., Tumlinson \& Shull 2000; Bromm,
Kudritzki, \& Loeb 2001b; Schaerer 2002, 2003; Tumlinson,
Shull, \& Venkatesan 2003).
The contribution from these stars, thus, could have been important
in reionizing the IGM early on, at $z\ga 15$, as may be required
by the recent {\it Wilkinson Microwave Anisotropy Probe (WMAP)} data
(Kogut et al. 2003; Spergel et al. 2003) on the optical depth to
Thomson scattering (e.g., Cen 2003a,b; Ciardi, Ferrara, \& White 2003;
Haiman \& Holder 2003; Holder et al. 2003; Kaplinghat et al. 2003;
Sokasian et al. 2003a,b; Wyithe \& Loeb 2003a,b). The second key feedback
effect concerns the production of the
first heavy elements, and their subsequent dispersal by supernova (SN)
explosions into the pristine IGM
(e.g., Gnedin \& Ostriker 1997; Madau, Ferrara, \& Rees 2001; Mori,
Ferrara, \& Madau 2002; Bromm, Yoshida, \& Hernquist 2003; Wada \&
Venkatesan 2003).
Once the IGM has been metal-enriched above a minimum level,
star formation is predicted to shift from a predominantly
high-mass (Population~III) to a lower-mass (Population~II) mode (e.g.,
Omukai 2000; Bromm et al. 2001a; Bromm \& Loeb 2003b; Schneider et al.
2003a). The nature of this fundamental transition, whether gradual
or sudden, will depend on the detailed IGM enrichment history
(e.g., Schneider et al. 2002; Mackey, Bromm, \& Hernquist 2003;
Scannapieco, Schneider, \& Ferrara 2003; Ricotti \& Ostriker 2004;
Yoshida, Bromm, \& Hernquist 2004).

Very recently, novel empirical probes have begun to test our theoretical
ideas on star and galaxy formation in the early universe. Next to the
{\it WMAP} data on the ionization state of the IGM, these probes include
determinations of the chemical abundance pattern in extremely metal-poor 
Galactic halo stars (e.g., Christlieb et al. 2002), measurements of the
near IR cosmic background radiation (e.g., Santos, Bromm, \& Kamionkowski
2002; Salvaterra \& Ferrara 2003), and the prospect of detecting
high-redshift gamma-ray bursts (GRBs) with the {\it Swift} satellite
to be launched in 2004 (e.g., Lamb \& Reichart 2000; Bromm \& Loeb 2002).

In this review, I will briefly summarize the current state of key aspects in
this rapidly evolving field (see
Bromm \& Larson 2004 for a more detailed discussion).

\section{BASIC COSMOLOGICAL FRAMEWORK}

The establishment of the current standard $\Lambda$CDM
model for cosmological structure formation has provided a firm framework
for the study of the first stars (Spergel et al. 2003). Within variants 
of the CDM model, where larger structures 
are assembled hierarchically through successive mergers of smaller 
building-blocks, the first stars are predicted to form in DM
minihalos of typical mass $\sim 10^{6} M_{\odot}$ at redshifts
$z\sim 20 - 30$ (e.g., Couchman \& Rees 1986). The virial temperatures
in these low-mass halos, $T_{\rm vir}\propto M^{2/3}(1+z)$ (Barkana \& 
Loeb 2001), are below the threshold, $\sim 10^{4}$~K, for efficient 
cooling due to atomic hydrogen (e.g.,
Oh \& Haiman 2002). It was realized
early on that cooling in the low-temperature
primordial gas had to rely on molecular hydrogen instead 
(Saslaw \& Zipoy 1967; Peebles \& Dicke 1968).

Since the thermodynamic behavior of the primordial gas thus
is primarily controlled by H$_{2}$ cooling, it is crucial to
understand the non-equilibrium chemistry of H$_{2}$ formation and destruction
(e.g., Lepp \& Shull 1984; Anninos \& Norman 1996; Abel et al. 1997;
Galli \& Palla 1998; Puy \& Signore 1999).
In the absence of dust grains to facilitate their formation (e.g., Hirashita
\& Ferrara 2002),
molecules have to form in the gas phase. The most important
formation channel is given by the sequence: H + $e^{-}\rightarrow$
H$^{-} + \gamma$, followed by H$^{-}+$ H $\rightarrow$ H$_{2}+ e^{-}$
(McDowell 1961). The free electrons act as catalysts, 
and are present as residue
from the epoch of recombination (Seager, Sasselov, \& Scott  2000),
or result from collisional ionization in accretion shocks
during the hierarchical build-up of galaxies
(e.g., Mac Low \& Shull 1986; Shapiro \& Kang 1987).
The formation of hydrogen molecules thus ceases when the free
electrons have recombined. 
Calculations of H$_{2}$ formation
in collapsing top-hat overdensities, idealizing the virialization
of dark matter halos in CDM cosmogonies, have found a simple
approximate relationship between the asymptotic H$_{2}$ abundance and
virial temperature in the overdensity (or halo): $f_{\rm H_{2}}
\propto T_{\rm vir}^{1.5}$ (Tegmark et al. 1997).

\begin{figure}[t]
\plotone{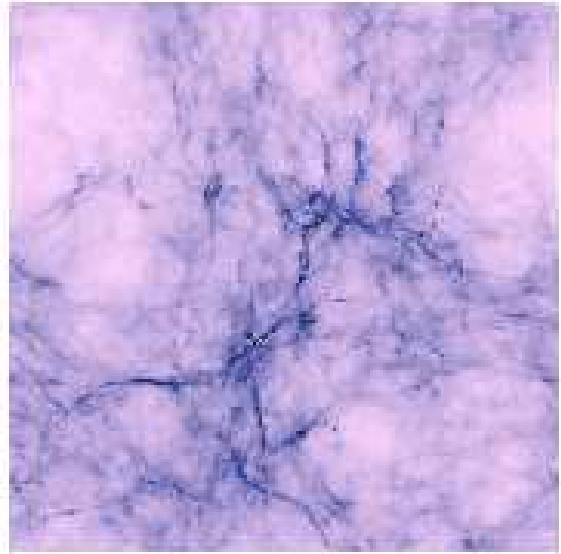}
\caption{Emergence of primordial star forming regions within
a standard $\Lambda$CDM cosmology. Shown is the projected gas density
at $z=17$ within a simulation box of physical size $\sim 50$~kpc. The bright
knots at the intersections of the filamentary network are the sites
where the first stars form.
(From Yoshida et al. 2003a.)}
\end{figure}

Applying the familiar criterion (Rees \& Ostriker 1977; Silk 1977)
for the formation of galaxies
that the cooling timescale has to be shorter than the dynamical
timescale, $t_{\rm cool} < t_{\rm dyn}$, one can derive
the minimum halo mass at a given redshift inside of which the gas
is able to cool and eventually form stars (e.g.,
Tegmark et al. 1997).
The H$_{2}$ cooling
function has been quite uncertain
over the relevant temperature regime, but
recent advances in the quantum-mechanical
computation of the collisional excitation process (H atoms colliding
with H$_{2}$ molecules) have provided a much more reliable
determination (see Galli \& Palla 1998, and
references therein).
Combining the CDM prescription for the assembly of virialized
DM halos with the H$_{2}$ driven thermal evolution of the primordial
gas, a minimum halo mass of $\sim 10^{6} M_{\odot}$
is required for collapse redshifts $z_{\rm vir}\sim 20-30$.
From detailed calculations, one finds that
the gas in such a `successful' halo has reached 
a molecule fraction in excess of
$f_{\rm H_{2}}\sim 10^{-4}$ (e.g., Haiman, Thoul, \& Loeb 1996;
Tegmark et al. 1997; Yoshida et al. 2003a).
These systems correspond to $3-4 \sigma$ peaks in the Gaussian
random field of primordial density fluctuations.
In principle, DM halos that are sufficiently massive to harbor
cold, dense gas clouds could form at higher redshifts, $z_{\rm vir}\ga 40$. Such systems, however,
would correspond to extremely rare, high-$\sigma$ peaks in the
Gaussian density field (e.g., Miralda-Escud\'{e} 2003).

To more realistically assess the formation of cold and dense star
forming clouds in the earliest stages of cosmological structure
formation, three-dimensional simulations of the combined
evolution of the DM and gas are required within a cosmological
set-up 
(e.g., Ostriker \& Gnedin 1996;
Gnedin \& Ostriker 1997; Abel et al. 1998).
These studies confirmed the important
role of H$_{2}$ cooling in low-mass halos at high $z$. Recently,
the problem of forming primordial gas clouds within a fully
cosmological context has been revisited with high numerical
resolution (Yoshida et al. 2003a,b,c). The resulting gas density
field is shown in Figure~1 for a standard $\Lambda$CDM cosmology at $z=17$.
The bright knots at the intersections of the filamentary network
are the star forming clouds, having individual masses (DM and gas)
of $\sim 10^{6} M_{\odot}$.

Whether a given DM halo successfully hosts a cold, dense ($T\la 0.5 
T_{\rm vir}$,$n_{\rm H}\ga 5\times 10^{2}$cm$^{-3}$) gas cloud
can nicely be understood with the Rees-Ostriker criterion, as can be
seen in Figure~2. Yoshida et al. (2003a) derive a minimum collapse
mass of $M_{\rm crit}\simeq 7\times 10^{5}M_{\odot}$, with only
a weak dependence on collapse redshift (see also Haiman et al. 1996;
Fuller \& Couchman 2000; Machacek, Bryan, \& Abel 2001). 
The dynamical heating accompanying
the merging of DM halos has an important effect on the thermal and
chemical evolution of the gas. Clouds do not successfully cool
if they experience too rapid a growth in mass.

The primordial gas clouds that are found in the cosmological
simulations are the sites where the first stars form. It is,
therefore, important to learn what properties these clouds have
in terms of overall size, shape, and angular momentum content.
The latter is often expressed by the familiar spin-parameter
$\lambda=L|E|^{1/2}/(G M^{5/2})$, where $L$, $E$, and $M$ are the
total angular momentum, energy, and mass, respectively. The spin
parameter is a measure of the degree of rotational support,
such that the ratio of centrifugal to gravitational acceleration
is given by $\sim \lambda^{2}$ at virialization. The spin values
measured in pure DM cosmological simulations
can be described by a lognormal distribution function with
a mean value of $\bar{\lambda}=0.04$, similar to what is found
for larger-scale systems (Jang-Condell \& Hernquist 2001).
The overall sizes of the Population~III
star forming clouds are close to the virial radius of the host
DM halo, with $R_{\rm vir}\sim 100$~pc, not too different
from the typical dimensions of present-day giant molecular clouds
(e.g., Larson 2003). Depending on the degree of spin, the clouds
have shapes with various degree of flattening (e.g., Bromm et al. 2002;
Yoshida et al. 2003a). Due to the importance of pressure forces, however,
overall cloud shapes tend to be rather spherical.
To fully elucidate the properties of the
star-forming primordial clouds, even higher resolution cosmological
simulations will be necessary.

\begin{figure}[t]
\plotone{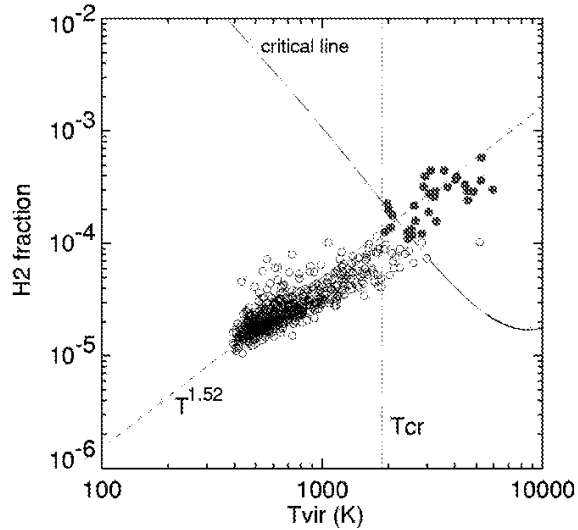}
\caption{Condition for successful formation of cold, dense gas
clouds. Mass-weighted mean H$_{2}$ fraction versus virial temperature.
{\it Open circles}: DM halos that fail to host cold gas clouds.
{\it Filled circles}: DM halos that succeed in harboring 
star-forming clouds. The critical line corresponds to the
condition $t_{\rm cool}\sim t_{\rm ff}$.
(From Yoshida et al. 2003a.)}
\end{figure}

The theoretical predictions for the formation sites of
the first stars sensitively depend on the exact nature of the
DM component and its fluctuation spectrum. Recently,
two models have been discussed that would much reduce the
fluctuation power on small mass scales. The first of these,
the warm dark matter (WDM) model (e.g., Bode, Ostriker, \& Turok 2001), 
has been proposed to remedy
the well-known problems of standard $\Lambda$CDM on sub-galactic
scales (e.g., Flores \& Primack 1994; Moore et al. 1999).
These concern the predicted large abundance of substructure or,
equivalently, of satellite systems, and the high (cuspy) 
densities in the centers of galaxies. Both predictions are
in conflict with observations.
The second model, the `running' spectral index (RSI) model,
is suggested by the combined analysis of the {\it WMAP} data,
the 2dF galaxy redshift survey, and Lyman-$\alpha$ forest observations
(Spergel et al. 2003; Peiris et al. 2003). A series of recent 
studies have worked out the consequences of these reduced small-scale
power models on early star formation (Somerville, Bullock, \& Livio 2003;
Yoshida et al. 2003b,c). 
Within these models, the star formation rate at $z\ga 15$ is significantly
reduced compared to the standard $\Lambda$CDM case. This is 
due to the absence of low-mass halos and their associated gas clouds that are
cooled by molecular hydrogen. Such a reduced rate of early star formation
makes it difficult to achieve the large optical depth to Thomson
scattering, as measured by
{\it WMAP}, with reasonable choices for the star formation efficiency in
the first galaxies.

\section{FORMING THE FIRST STARS}

\subsection{Formation of Prestellar Clumps}

The metal-rich chemistry, magnetohydrodynamics, and radiative
transfer involved in present-day star formation are complex, and we
still lack a comprehensive theoretical framework that predicts the IMF
from first principles. Star formation in the high redshift universe,
on the other hand, poses a theoretically more tractable problem due to
a number of simplifying features, such as: (i) the initial absence of
heavy elements and therefore of dust; (ii) the absence of
dynamically-significant primordial magnetic fields; 
and (iii) the absence of any effects from
previous episodes of star formation which would completely alter
the conditions for subsequent generations of stars.
The cooling of the primordial gas depends only
on hydrogen in its atomic and molecular form.  Whereas in the
present-day interstellar medium (ISM) the initial state of the star forming
cloud is poorly constrained, the corresponding initial conditions for
primordial star formation are simple, given by the popular
$\Lambda$CDM model of cosmological structure formation.

How did the first stars form? This subject has a long and venerable
history (e.g., Schwarzschild \& Spitzer 1953;
Yoneyama 1972; Hutchins 1976; Silk 1977, 1983; Yoshii \& Sabano 1979;
Carlberg 1981; Kashlinsky \& Rees 1983;
Palla, Salpeter, \& Stahler 1983; Yoshii \& Saio 1986). In this review, I focus
mainly on the more recent work since the renewed interest in high-redshift
star formation beginning in the mid-1990s (e.g., Haiman et al. 1996;
Uehara et al. 1996; Haiman \& Loeb 1997; Tegmark et al. 1997; Larson 1998).
The complete answer to this question
would entail a theoretical prediction for the Population~III IMF,
which is rather challenging. A more tractable task is to estimate
the characteristic mass scale, $M_c$, of the first stars, and most
of the recent numerical work has focused on this simpler problem.
The characteristic mass is the mass below which the IMF flattens
or begins to decline (see Larson 1998 for examples of possible
analytic forms).
This mass scale is observed to be $\sim 1
M_{\odot}$ in the present-day universe. 
Since the detailed shape of the primordial IMF is highly uncertain,
it is reasonable to first constrain $M_c$, as this mass scale indicates
the typical outcome of the primordial star formation process or, in other words,
where most of the available mass ends up (e.g., Clarke \& Bromm 2003).

To fully explore the dynamical, thermal, and chemical properties of primordial
gas, three-dimensional numerical simulations are needed, although more
idealized investigations in two, one, or even zero 
dimension are important in that they allow to probe a larger parameter space.
The proper initial conditions for primordial star formation are given
by the underlying model of cosmological structure formation (see \S~2).
It is, therefore, necessary to simulate both the DM and
gaseous (`baryonic') components. To date, two different numerical approaches
have been used to simulate the general three-dimensional fragmentation
problem in its cosmological context. The first series of simulations
used the grid-based adaptive mesh-refinement (AMR) technique
(Abel et al. 2000, 2002; ABN henceforth), and the second one employed
the smoothed particle hydrodynamics (SPH)
method (Bromm et al. 1999, 2002; BCL henceforth). 
The SPH approach
has the important advantage that it can easily accommodate
the creation of sink particles (e.g., Bate, Bonnell, \& Bromm 2003).
Recently, the dynamical range of the standard SPH method has been
significantly improved by implementing a `particle splitting'
technique (Kitsionas \& Whitworth 2002; Bromm \& Loeb 2003a).
When the simulation reaches such high density in a certain region
that a sink particle would normally be created, a complementary strategy
is now adopted: Every SPH particle in the unrefined, high-density
region acts as a parent particle and spawns a given number of child particles,
and endows them with its properties. In effect, this is the SPH equivalent
of the grid-based AMR technique.

The most important difference between these two studies lies in the way
the initial conditions are implemented. The ABN simulations start at
$z=100$ with a periodic
volume of physical size $128/(1+z)$~kpc. The AMR technique allows ABN
to bridge the gap from cosmological to protostellar scales. The BCL effort,
on the other hand, initializes the simulations, also at $z=100$, 
by realizing
spherical overdensities that correspond to
high-$\sigma$ peaks in the Gaussian random field of cosmological density
fluctuations.

\begin{figure}[t]
\plotone{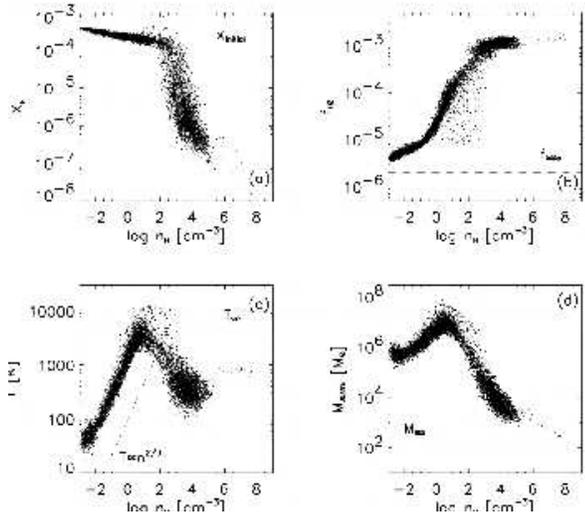}
\caption{
Properties of primordial gas.
{\bf (a)} Free electron abundance vs. hydrogen number density (in cm$^{-3}$).
{\bf (b)} Hydrogen molecule abundance vs. number density. 
{\bf (c)} Gas temperature vs. number density. At densities below $\sim 1$ cm$^
{-3}$, the gas temperature rises because of adiabatic compression until
it reaches the virial value of $T_{vir}\simeq 5000$ K.
At higher densities, cooling due to H$_{2}$
drives the temperature down again, until the gas settles into a quasi-
hydrostatic state at $T\sim 200$ K and $n\sim 10^{4}$ cm$^{-3}$.
Upon further compression due to the onset of the gravitational
instability, the temperature experiences a modest rise again.
{\bf (d)} Jeans mass (in $M_{\odot}$) vs. number density. The Jeans mass
reaches a value of $M_{J}\sim 10^{3}M_{\odot}$ for the quasi-hydrostatic
gas in the center of the DM potential well.
(From Bromm et al.\ 2002.)}
\end{figure}

In comparing the simulations of ABN and BCL, the most important aspect
is that both studies, employing very different methods, agree on the
existence of a preferred state for the primordial gas, corresponding
to characteristic values of temperature and density,
$T_c\sim 200$~K, and
$n_c\sim 10^{4}$ cm$^{-3}$, respectively. These characteristic scales
turn out to be rather robust in the sense that they are not very
sensitive to variations in the initial conditions.
In Figure~3 (panel c), this preferred state in the $T-n$ phase diagram
can clearly be discerned. This figure, from the simulations of BCL,
plots the respective gas properties for each individual SPH particle.
The diagram thus contains an additional dimension of information:
where evolutionary timescales are short, only few particles are
plotted, whereas they tend to accumulate where the overall evolution is slow.
Such a `loitering' state is reached at $T_c$ and $n_c$.

These characteristic scales can be understood by considering
the microphysics of H$_2$, the main coolant in metal-free, star forming
gas (Abel et al. 2002; Bromm et al. 2002). At temperatures $\la 1000$~K,
cooling is due to the collisional excitation and subsequent radiative 
decay
of rotational transitions. The two lowest-lying rotational energy levels
in H$_2$ have an energy spacing of $E/k_{\rm B}\simeq 512{\rm ~K}$. 
Collisions with particles (mostly H atoms) that populate the high-energy
tail of the Maxwell-Boltzmann velocity distribution can lead to somewhat
lower temperatures, but H$_2$ cooling cannot proceed to $T\la 100$~K.
This explains the characteristic temperature. The characteristic density,
in turn, is given by the critical density above which collisional
de-excitations, which do not cool the gas,  compete with 
radiative decays, which lead to cooling. This saturation
of the H$_2$ cooling marks the transition from NLTE rotational
level populations to thermal (LTE) ones. At densities below $n_c$, the
cooling rate is proportional to the density squared, whereas at higher
densities, the dependence is only linear.

Once the characteristic state is reached, the evolution towards higher
density is temporarily halted due to the now inefficient cooling, and the
gas undergoes a phase of quasi-hydrostatic, slow contraction.
To move away from this `loitering' regime, enough mass has to accumulate
to trigger a gravitational runaway collapse. This condition is simply
$M \ga M_J$, where the Jeans mass or, almost equivalently, the
Bonnor-Ebert mass can be written as (e.g., Clarke \& Bromm 2003)
\begin{equation}
M_J\simeq 700 M_{\odot}\left(\frac{T}{200{\rm ~K}}\right)^{3/2}
\left(\frac{n}{10^{4}{\rm cm}^{-3}}\right)^{-1/2} \mbox{\ .}
\end{equation}
Primordial star formation, where magnetic fields and turbulence are expected
to initially play no important dynamical role, may be the best case
for the application of the classical Jeans criterion which is describing
the balance between gravity and the opposing {\it thermal} pressure.

A prestellar clump of mass $M\ga M_J$ is the immediate progenitor
of a single star or, in case of further subfragmentation, a binary
or small multiple system. In Galactic star-forming regions, like
$\rho$~Ophiuchi, such clumps with masses close to stellar values
have been observed as gravitationally bound clouds which lack the
emission from embedded stellar sources (e.g., Motte, Andr\'{e}, \& Neri 1998).
The high-density clumps are clearly not stars yet. To probe the
further fate of a clump, one first has to follow the collapse to
higher densities up to the formation of an optically thick hydrostatic
core in its center (see \S~3.2), and subsequently the accretion from
the diffuse envelope onto the central core (see \S~3.3). The parent
clump mass, however, already sets an upper limit for the final stellar
mass whose precise value is determined by the accretion process.

Although ABN and BCL agree on the magnitude of the characteristic
fragmentation scale, ABN have argued that only one star forms per halo,
while BCL have simulated cases where multiple clumps
form, such that the number of stars in a minihalo is $N_{\ast}\sim 1 - 5$.
Among the cases studied by BCL favoring $N_{\ast} > 1$, are high-spin runs that
lead to disk-like configurations which subsequently fragment into
a number of clumps, or more massive halos (with total mass $M \la 10^{7}
M_{\odot}$). Notice that the case simulated by ABN corresponds to
a gas mass of $\sim 3\times 10^{4} M_{\odot}$. Such a low-mass halo is
only marginally able to cool, and when BCL simulated a system with this mass,
they also found that only a single star forms inside the halo.
Further work is required to elucidate whether the multiple-clump formation
in BCL does also occur in simulations with realistic cosmological
initial conditions.

A possible clustered nature of forming Population~III stars would have
important consequences for the transport of angular momentum. In fact,
a crucial question is how the primordial gas can so efficiently shed
its angular momentum on the way to forming massive clumps.
In the case of a clustered formation process, angular momentum could
be efficiently transported outward by gravitational (tidal) torques.
Tidal torques can then transfer much of the angular momentum from
the gas around each forming clump to the orbital motion of the system,
similar to the case of present-day star formation in a clustered
environment (e.g., Larson 2002; Bate et al. 2003). In the isolated
formation process of ABN, however, such a mechanism is not available.
Alternatively, ABN have suggested the efficient transport of
angular momentum via hydrodynamic shocks during turbulent collapse.
Again, more work is required to convincingly sort out the angular momentum
transport issue. 

Important contributions have also been made
by more idealized, two- to zero-dimensional, studies. 
These investigations typically ignore the dark matter component, thus
implicitly assuming that the gas has already dissipatively collapsed
to the point where the dark matter ceases to be dynamically dominant.
Much attention has been paid to the collapse of filamentary clouds
(Uehara et al. 1996; Nakamura \& Umemura 1999, 2001, 2002).
Results from one- and two-dimensional simulations that otherwise 
include all the relevant processes for the primordial chemistry and
cooling have estimated fragmentation scales $M_c \sim 1 M_{\odot}$
and $M_c \sim 100 M_{\odot}$, depending on the (central) initial density.
When sufficiently high ($n \ga 10^5$cm$^{-3}$), the low-mass value is reached.
This bifurcation has led to the prediction of a bimodal IMF for the first
stars (Nakamura \& Umemura 2001). It is however, not obvious how
such a high initial density can be reached in a realistic situation
where the collapse starts from densities that are typically much smaller
than the bifurcation value.

\subsection{Protostellar collapse}

What is the further fate of the clumps discussed above?
In particular, one would like to test the notion that such a Jeans unstable
clump is the immediate progenitor of a single star which forms in its center.
That will evidently only be correct if the clump does not undergo further
subfragmentation upon collapsing to higher densities.
It has long been suspected that such subfragmentation could occur at
densities in excess of $\sim 10^{8}$cm$^{-3}$, at which point three-body
reactions become very efficient in converting the atomic gas (with only
a trace amount of H$_2$ molecules from the H$^-$ channel) into fully
molecular form: 3H $\rightarrow$ H$_2$ + H (Palla et al. 1983).
As the H$_2$ coolant is now suddenly more abundant by a factor of $\sim 10^{3}$,
the corresponding boost in cooling could trigger a thermal instability,
thus breaking up the clump into smaller pieces (Silk 1983).
Both ABN and BCL have included the three-body reactions in their chemical 
reaction networks, and have followed their simulations to higher densities
to test whether subfragmentation does occur or not.
Both groups report that no
further subfragmentation is seen. With hindsight, that may not be too
surprising. The reason being that any small density fluctuations which
are present earlier on, and which could serve as seeds for later fragmentation,
will have been erased by pressure forces during the slow, quasi-hydrostatic
`loitering' phase at $n \sim n_c$. 
In addition, inefficient cooling may also play a role in suppressing
high-density fragmentation. Despite the increase in the cooling rate 
throughout the fully molecular gas, this never
leads to a significant drop in temperature due to the countervailing effect
of compressional heating.

\begin{figure}[t]
\plotone{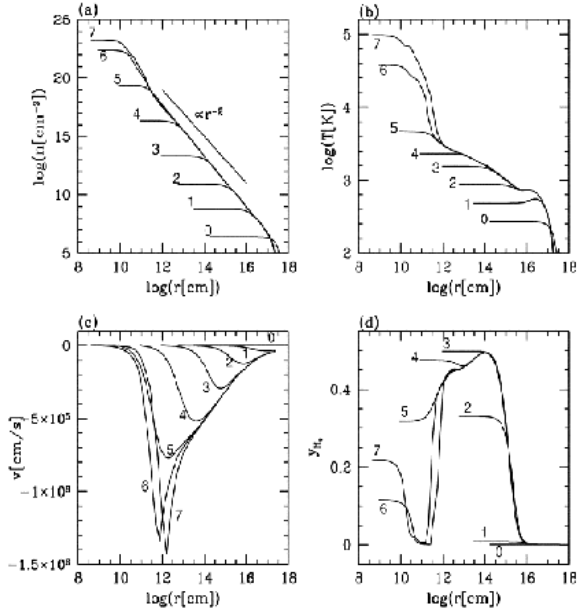}
\caption{
Collapse of a primordial protostar. Shown is the evolutionary sequence
as obtained in a 1D calculation, with time progressing from curves
labeled 1 to 7.
{\bf (a)} Number density vs. radial distance.
{\bf (b)} Gas temperature vs. distance. 
{\bf (c)} Radial velocity vs. distance.
{\bf (d)} Hydrogen molecule abundance vs. distance.
The curves labeled 6 correspond to the situation briefly after the
formation of a central hydrostatic core.
(From Omukai \& Nishi 1998.)}
\end{figure}

Extending the analogous calculation for the collapse of a present-day
protostar (Larson 1969) to the primordial case, Omukai \& Nishi (1998)
have carried out one-dimensional hydrodynamical simulations in spherical
symmetry. They also consider the full set of chemical reactions, and
implement an algorithm to solve for the radiative transfer in the
H$_2$ lines, as well as in the continuum. The most important result is that
the mass of the initial hydrostatic core, formed in the center of the
collapsing cloud when the density is high enough ($n\sim 10^{22}$cm$^{-3}$)
for the gas to become optically thick to continuum radiation, is almost
the same as the (second) core in present-day star formation: $M_{\rm core}
\sim 5\times 10^{-3} M_{\odot}$. 
In Figure~4, the radial profiles
of density, temperature, velocity, and H$_2$ abundance are shown (reproduced
from Omukai \& Nishi 1998). The profiles of density and velocity
before the time of core formation (corresponding to the curves with label 6)
are well described by the Larson-Penston (LP) similarity solution. Once the
core has formed, the self-similarity is broken.
Similar results have been found by Ripamonti et al. (2002) who have 
in addition worked out the spectrum of the radiation that escapes
from the collapsing clump (mostly in the IR, both as continuum and line
photons). This type of approximately self-similar behavior seems to be
a very generic result of collapse with a simple
equation of state, even when rotation and magnetic fields are 
included (see Larson 2003). The apparent robustness of the LP solution
thus supports the results found for the Population~III case.
The small value of the initial core does not mean that this will be
the final mass of a Population~III star. This instead will be determined by
how efficient the accretion process will be in incorporating the clump mass
into the growing protostar. 

\subsection{Accretion Physics}

\begin{figure}[t]
\plotone{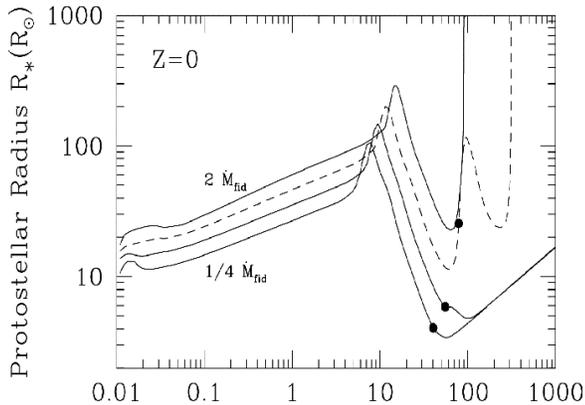}
\caption{
Evolution of accreting metal-free protostar.
Shown is the radius-mass relation for different values of the
accretion rate (increasing from bottom to top). Accretion is
effectively shut off for the cases with $\dot{M}\ga \dot{M}_{\rm crit}$
because of the dramatic increase in radius.
(From Omukai \& Palla 2003.)}
\end{figure}

How massive were the first stars? Star formation typically
proceeds from the `inside-out', through the accretion of gas onto a
central hydrostatic core.  Whereas the initial mass of the hydrostatic
core is very similar for primordial and present-day star formation
(see above), the accretion process -- ultimately responsible for
setting the final stellar mass -- is expected to be rather different. On
dimensional grounds, the accretion rate is simply related to the sound
speed cubed over Newton's constant (or equivalently given by the ratio
of the Jeans mass and the free-fall time): $\dot{M}_{\rm acc}\sim
c_s^3/G \propto T^{3/2}$. A simple comparison of the temperatures in
present-day star forming regions ($T\sim 10$~K) with those in
primordial ones ($T\sim 200-300$~K) already indicates a difference in
the accretion rate of more than two orders of magnitude.

Recently, Omukai \& Palla (2001, 2003)
have investigated the accretion problem in considerable detail,
going beyond the simple dimensional argument given above.
Their computational technique approximates the time evolution
by considering a sequence of steady-state accretion flows onto a
growing hydrostatic core. Somewhat counterintuitively, these authors
identify a critical accretion rate, $\dot{M}_{\rm crit}\sim 4\times 10^{-3}
M_{\odot}$~yr$^{-1}$, such that for accretion rates higher than this,
the protostar cannot grow to masses much in excess of a few $10 M_{\odot}$.
For smaller rates, however, the accretion is predicted to
proceed all the way up to $\sim 600 M_{\odot}$, i.e., of order the
host clump.

The physical basis for the critical accretion rate is that
for ongoing accretion onto the core, the luminosity must not exceed
the Eddington value, $L_{\rm EDD}$. In the early stages of accretion,
before the onset of hydrogen burning, the luminosity is approximately 
given by $L_{\rm tot}\sim L_{\rm acc}\simeq G M_{\ast}\dot{M}_{\rm acc}/R_{\ast}$.
By demanding $L_{\rm acc}\simeq L_{\rm EDD}$ it follows
\begin{equation}
\dot{M}_{\rm crit}\simeq \frac{L_{\rm EDD}R_{\ast}}{G M_{\ast}}
\sim 5\times 10^{-3} M_{\odot}{\rm ~yr}^{-1} \mbox{\ ,}
\end{equation}
where $R_{\ast}\sim 5 R_{\odot}$, a typical value for a Population~III main-sequence
star (e.g., Bromm et al. 2001b). In Figure~5 (from Omukai \& Palla 2003),
the mass-radius relation is shown for various values of the
accretion rate. As can be seen, the dramatic swelling in radius
effectively shuts off  accretion at $M_{\ast}\la 100 M_{\odot}$, when
the accretion rate exceeds $M_{\rm crit}$.

Realistic accretion flows are expected to have a time-dependent rate,
and the outcome will thus depend on whether the accretion
rate will decline rapidly enough to avoid exceeding the Eddington
luminosity at some stage during the evolution.
The biggest caveat concerning the Omukai \& Palla results
seems to be the issue of geometry. A three-dimensional accretion
flow of gas with some residual degree of angular momentum
will deviate from spherical symmetry, and instead form a disk. It is
then conceivable that most of the photons can escape along the axes
whereas mass can flow in unimpeded through the accretion disk (see
Tan \& McKee 2003).

As described above, the accretion process may be able to incorporate
a large part of the parent clump into the central star.
{\it Can a Population~III star ever reach this asymptotic mass limit?}
The answer to this question is not yet known with any certainty, and
it depends on whether the accretion from the dust-free envelope is eventually
terminated by feedback from the star (e.g., Omukai \& Palla 2001, 2003;
Omukai \& Inutsuka 2002; Ripamonti et al. 2002; Tan \& McKee 2003).
The standard mechanism by which
accretion may be terminated in metal-rich gas, namely radiation pressure
on dust grains (Wolfire \& Cassinelli 1987), is evidently not effective for
gas with a primordial composition. Recently, it has been speculated
that accretion could instead be turned off through the formation of an
\ion{H}{2} region (Omukai \& Inutsuka 2002), or through the radiation pressure exerted
by trapped Ly$\alpha$ photons (Tan \& McKee 2003). The termination of the
accretion process defines the current unsolved frontier in studies of
Population~III star formation.

\subsection{The Second Generation of Stars: Critical Metallicity}

How and when did the transition take place from the
early formation of massive stars to that of low-mass stars at later
times?  
The second generation of stars formed under conditions that
were much more complicated
again due to the feedback from the first stars on the IGM, both
due to the production of photons and of heavy elements.
In contrast to the formation mode of
massive stars (Population~III) at high redshifts, fragmentation is observed
to favor stars below a solar mass (Population~I and II) in the present-day
universe.  The transition between these fundamental modes is expected to be
mainly driven by the progressive enrichment of the cosmic gas with heavy
elements, which enables the gas to cool to lower temperatures.
The concept of a `critical metallicity', $Z_{\rm crit}$, has been
used to characterize the transition between
Population~III and Population~II formation modes, where $Z$ denotes the
mass fraction contributed by all heavy elements (Omukai 2000;
Bromm et al. 2001a; Schneider et al. 2002, 2003a; Mackey et al. 2003).
These studies have constrained
this important parameter to only within a few orders of magnitude, $Z_{\rm
crit}\sim 10^{-6}-10^{-3} Z_{\odot}$, under the implicit assumption of
solar relative abundances of metals.
This assumption is likely to be
violated by the metal yields of the first SNe at
high-redshifts, for which strong deviations from solar abundance ratios are
predicted (e.g., Oh et al. 2001; Heger \& Woosley 2002; 
Qian, Sargent, \& Wasserburg 2002; Qian \& 
Wasserburg 2002; Umeda \& Nomoto 2002, 2003).

Recently, Bromm \& Loeb (2003b)
have shown that the transition between the above star formation modes is
driven primarily by fine-structure line cooling of singly-ionized carbon or
neutral atomic oxygen.  Earlier estimates of $Z_{\rm crit}$ which
did not explicitly distinguish between different coolants are refined
by introducing
separate critical abundances for carbon and oxygen, [C/H]$_{\rm crit}$ and
[O/H]$_{\rm crit}$, respectively, where [A/H]= $\log_{10}(N_{\rm A}/N_{\rm
H})-\log_{10}(N_{\rm A}/N_{\rm H})_{\odot}$.
Since C and O are also the most important coolants
throughout most of the cool atomic ISM in present-day
galaxies, it is not implausible that these species might be
responsible for the global shift in the star formation mode.
Under the temperature and density conditions that characterize
Population~III star formation, the fine-structure lines of \ion{O}{1} and
\ion{C}{2} dominate over all other metal
transitions (see Hollenbach \& McKee 1989). 
Cooling due to molecules becomes important only at
lower temperatures, and cooling due to dust grains only at higher
densities (e.g., Omukai 2000; Schneider et al. 2003a). The presence of
dust is likely to modify the equation of state at
these high densities in important ways (Schneider et al. 2003a), and it will
be interesting to explore its role in future collapse calculations.
The physical nature of the dust, however, that is produced by the first SNe
is currently still quite uncertain (e.g., Loeb \& Haiman 1997;
Todini \& Ferrara 2001; Nozawa et al. 2003;
Schneider, Ferrara, \& Salvaterra 2003b).
Numerically, the critical C and O abundances are estimated to be:
[C/H]$_{\rm
crit}\simeq -3.5 \pm 0.1$ and
[O/H]$_{\rm crit}\simeq -3.1 \pm 0.2$.

Even if sufficient C or O atoms are present to further cool the
gas, there will be a minimum attainable temperature
that is set by the interaction of the atoms with the thermal CMB: $T_{\rm
CMB}=2.7{\rm \,K}(1+z)$ (e.g., Larson 1998; Clarke \& Bromm 2003).
At $z\simeq 15$, this results in a characteristic
stellar mass of $M_{\ast}\sim 20 M_{\odot}(n_f/ 10^{4}{\rm
\,cm}^{-3})^{-1/2}$, where $n_f> 10^{4}{\rm \,cm}^{-3}$ is the density at
which opacity prevents further fragmentation (e.g., Rees 1976). 
It is possible that the transition from the high-mass to the low-mass
star formation mode was modulated by the CMB temperature and was therefore
gradual, involving intermediate-mass (`Population~II.5') stars
at intermediate redshifts (Mackey et al. 2003).
This transitional population could give rise to
the faint SNe that have been proposed to explain the observed abundance
patterns in metal-poor stars (Umeda \& Nomoto 2002, 2003).
When and how uniformly the transition in the cosmic star formation
mode did take place was governed by the detailed enrichment history
of the IGM. This in turn was determined by the hydrodynamical
transport and mixing of metals from the first SN explosions (e.g.,
Mori et al. 2002; Bromm et al. 2003; Scannapieco et al. 2003; Wada
\& Venkatesan 2003). The transport and the mixing of the first heavy elements
into the pristine IGM is currently another subject at the frontier
of astrophysical cosmology.

\subsection{Stellar Archaeology: Relics from the End of the Dark Ages}

It has long been realized that the most metal-poor stars found in
our cosmic neighborhood would encode the signature from the first stars
within their elemental abundance pattern
(e.g., Bond 1981; Beers, Preston, \& Shectman 1992).
For many decades, however, the observational search has failed
to discover a truly first-generation star with zero metallicity. Indeed,
there seemed to have been an observational lower limit of [Fe/H] $\sim -4$ 
(e.g., Carr 1987).
In view of the recent theoretical prediction that most Population~III stars
were very massive, with associated lifetimes of $\sim 10^6$~yr, the failure
to find any `living' Population~III star in the Galaxy is not surprising,
as they would all have died a long time ago (e.g., Hernandez \& Ferrara 2001).
Furthermore, theory has
predicted that star formation out of extremely low-metallicity gas,
with $Z\la Z_{\rm crit}\sim 10^{-3.5}Z_{\odot}$ (see \S~3.4), would
be essentially equivalent to that out of truly primordial gas.
Again, this theoretical prediction was in accordance with the apparent
observed lower-metallicity cutoff.

\begin{figure}[t]
\plotone{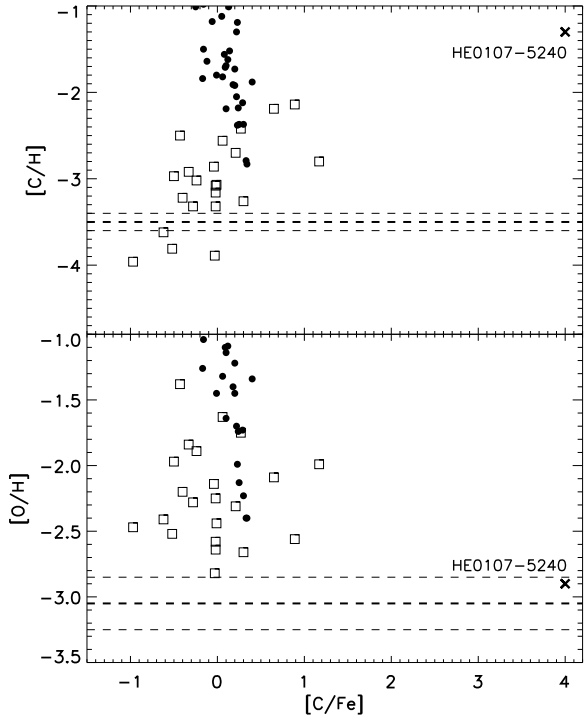}
\caption{Observed abundances in low-metallicity Galactic halo stars.
For both carbon ({\it upper panel}) and oxygen ({\it lower panel}),
filled circles correspond to samples of dwarf and subgiant stars
(from Akerman et al. 2003), and open squares to a sample of
giant stars (from Cayrel et al. 2003). The dashed lines indicate
the predicted critical carbon and oxygen abundances (see \S~3.4).
Highlighted ({\it cross}) is the location
of the extremely iron-poor  giant star HE0107-5240.
(Adapted from Bromm \& Loeb 2003b.)}
\label{fig6}
\end{figure}

Recently, however, this simple picture has been challenged by the
discovery of the star HE0107-5240 with a mass of $0.8 M_{\odot}$
and an {\it iron} abundance of ${\rm [Fe/H]} = -5.3$ 
(Christlieb et al. 2002). This finding indicates that at least
some low mass stars could have formed out of extremely low-metallicity gas.
Does the existence of this star invalidate the theory of a metallicity
threshold for enabling low-mass star formation? As pointed out 
by Umeda \& Nomoto (2003), a possible explanation could lie in
the unusually high abundances of carbon and oxygen in HE0107-5240.

In Figure~6, the theoretical C and O thresholds derived by Bromm \& Loeb
(2003b) are compared to the observed 
abundances in metal-poor dwarf (Akerman et al. 2003) 
and giant (Cayrel et al. 2003) stars in
the halo of our Galaxy.  As can be seen, all data points lie above the
critical O abundance but a few cases lie below the critical C threshold.
All of these low mass stars are consistent with the model since the
corresponding O abundances lie above the predicted threshold.  The
sub-critical [C/H] abundances could have either originated in the
progenitor cloud or from the mixing of CNO-processed material (with
carbon converted into nitrogen) into the stellar atmosphere during the red
giant phase. 
Note that the extremely iron-poor
star HE0107-5240 has C and O abundances that both lie above the respective
critical levels. The formation of this low mass star ($\sim 0.8 M_{\odot}$)
is therefore consistent with the theoretical framework considered
by Bromm \& Loeb (2003b).

The lessons from stellar archaeology on the nature of the first stars
are likely to increase in importance, since greatly improved, large
surveys of metal-poor Galactic halo stars are under way, or are currently
being planned.

\section{THE FIRST QUASARS}

Quasars are believed to be powered by the heat generated during the
accretion of gas onto SMBHs (e.g.,
Rees 1984).  The existence of SMBHs with inferred masses of
$\ga 10^{9}M_{\odot}$, less than a billion years after the big bang,
as implied by the recent discovery of quasars at redshifs $z\ga 6$
(e.g., Fan et al. 2003),
provides important
constraints on the SMBH formation scenario (Haiman \& Loeb 2001).  
{\it Can the seeds
of SMBHs form through the direct collapse of primordial gas clouds at
high redshifts?} Previous work (Loeb \& Rasio 1994) has shown that without a
pre-existing central point mass, this is rendered difficult by the
negative feedback resulting from star formation in the collapsing
cloud. The input of kinetic energy due to supernova explosions
prevents the gas from assembling in the center of the dark matter
potential well, thus precluding the direct formation of an SMBH. If,
however, star formation was suppressed in a cloud that could still
undergo overall collapse, such an adverse feedback would not occur.

\begin{figure}[t]
\plotone{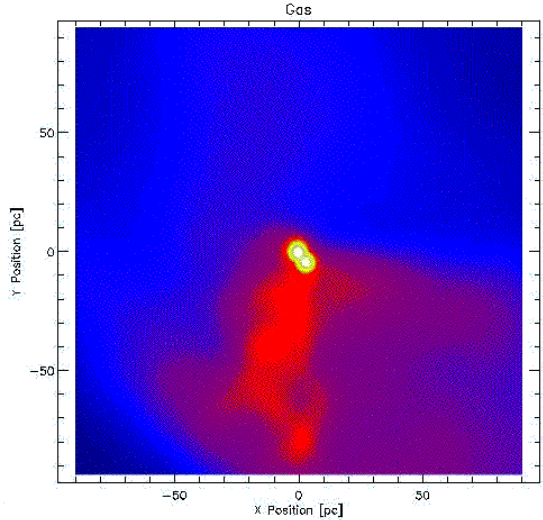}
\caption{Central gas density in a dwarf galaxy with virial temperature
just above the atomic cooling threshold, but with no H$_{2}$ molecules
present.
Shown is the projection in the $x-y$ plane at $z\sim 10$. 
The box size is 200~pc on a
side.  
In this case, the initial spin is
$\lambda=0.05$. Two compact objects have formed with masses
of $2.2 \times 10^{6}M_{\odot}$ and $3.1 \times 10^{6}M_{\odot}$,
respectively, and radii $\la 1$~pc.
(Adapted from Bromm \& Loeb 2003a).} 
\end{figure}

The SMBH formation problem has recently been revisited with
SPH simulations of isolated
2$\sigma-$peaks with total masses of $10^{8}M_{\odot}$ that collapse
at $z_{\rm vir}\sim 10$ (Bromm \& Loeb 2003a).
The virial temperature of these dwarf
galaxies exceeds $\sim 10^4$K, so as to allow collapse of their gas
through cooling by atomic hydrogen transitions (Oh \& Haiman 2002). Since
structure formation proceeds in a bottom-up fashion, such a system
would encompass lower-mass halos that would have collapsed earlier
on. These sub-systems have virial temperatures below $10^{4}$~K, and
consequently rely on the presence of H$_{2}$ for their
cooling. Molecular hydrogen, however, is fragile and readily destroyed
by photons in the Lyman-Werner bands with energies
($11.2-13.6$~eV) just below the Lyman limit (Haiman, Rees, \& Loeb 1997). These
photons are able to penetrate a predominantly neutral IGM.

At first, the limiting case is considered where H$_{2}$ destruction is
complete.  Depending on the initial spin, 
either one
(for zero initial spin) or two compact objects form with masses in
excess of $10^{6}M_{\odot}$ and radii $< 1$~pc (see Fig.~7). In the
case of nonzero spin a binary system of clumps has formed with a
separation of $\sim 10$~pc. Such a system of two compact objects is
expected to efficiently radiate gravitational waves that could be
detectable with the planned {\it Laser Interferometer Space
Antenna}\footnote{See http://lisa.jpl.nasa.gov/} (LISA; Wyithe \& Loeb 2003c).

What is the further fate of the central object? Once the gas has
collapsed to densities above $\sim 10^{17}$~cm$^{-3}$ and radii
$<10^{16}$~cm, Thomson scattering traps the photons, and the cooling
time consequently becomes much larger than both the free-fall and
viscous timescales (see Bromm \& Loeb 2003a for details). The gas is
therefore likely to settle into a radiation-pressure supported
configuration resembling a rotating supermassive star. Recent
fully-relativistic calculations of the evolution of such stars predict
that they would inevitably collapse to a massive black hole
(Baumgarte \& Shapiro 1999; Shibata \& Shapiro 2002).
Under a wide range of initial conditions, a
substantial fraction ($\sim 90$\%) of the mass of the supermassive
star is expected to end up in the black hole.

Is such a complete destruction of H$_{2}$ possible? When including an
external background of soft UV radiation 
in the simulation, Bromm \& Loeb (2003a) find
that a flux level comparable to what is expected close to the end of
the reionization epoch is sufficient to suppress H$_2$ molecule
formation. This is the case even when the effect of self-shielding is taken
into account (Draine \& Bertoldi 1996). The effective suppression of H$_{2}$
formation crucially depends on the presence of a stellar-like
radiation background. It is therefore likely that stars existed before
the first quasars could have formed.

\section{CONCLUSIONS}

The first stars and quasars, their formation and their cosmological
implications, are a fascinating subject at the frontier of modern
cosmology. Up to now, most of our efforts have been theoretical, but
we are approaching the point where observations can test our ideas.
This is very significant, as we are bound to learn important lessons about
the physical state of the universe at the end of its dark age.
These empirical probes are provided by both high-redshift observations,
and complementary to this, by the `near-field cosmology' (Freeman \&
Bland-Hawthorn 2002) of scrutinizing the chemical abundance patterns
in extremely metal-poor stars in our local cosmic neighborhood.
One can only wonder what exciting discoveries are awaiting us, and 
it is a great privilege to be a part of this grand endeavor.

\acknowledgments

I would like to thank my Ph.D. thesis advisers Paolo Coppi and
Richard Larson for all their help and support, as well as 
Yale's Department of Astronomy for providing me with an excellent
research environment.
I am grateful to C. Clarke, P. Demarque, A. Ferrara,
L. Hernquist, M. Kamionkowski, A. Loeb, J. Mackey,
M. Santos, J. Wasserburg, and N. Yoshida for the many discussions on the subject
discussed here. I thank Avi Loeb and Lars Hernquist at the 
Harvard-Smithsonian Center for Astrophysics for support from NSF
grant AST 00-71019 and NASA grant NAG5-13292,
as well as Cathie Clarke at the Institute
of Astronomy in Cambridge from the European Community's Research
Training Network under contract HPRN-CT-2000-0155,`Young Stellar
Clusters' and the Leverhulme Trust.




\begin{thebibliography}{}
\bibitem[aanz98]{aanz98} Abel, T., Anninos, P., Norman, M. L., \& Zhang, Y. 
1998, \apj, 508, 518
\bibitem[aazn97]{aazn97} Abel, T., Anninos, P., Zhang, Y., \& Norman, M. L. 
1997, NewA, 2, 181
\bibitem[abn00]{abn00} Abel, T., Bryan, G., \& Norman, M. L. 2000, 
\apj, 540, 39
\bibitem[abn02]{abn02} Abel, T., Bryan, G., \& Norman, M. L. 2002, 
Science, 295, 93
\bibitem[]{ak03} Akerman, C. J., Carigi, L., Nissen, P. E., 
Pettini, M., \& Asplund, M. 2003, \aap, in press (astro-ph/0310472)
\bibitem[an96]{an96} Anninos, P., \& Norman, M. L. 1996, \apj, 460, 556 
\bibitem[bl01]{bl01} Barkana, R., \& Loeb, A. 2001, Phys. Rep., 349, 125 
\bibitem[bbb03]{bbb03} Bate, M. R., Bonnell, I. A., \& Bromm, V. 2003, 
\mnras, 339, 577
\bibitem[]{bs99} Baumgarte, T. W., \& Shapiro, S. L. 1999, \apj, 526, 941
\bibitem[BPS]{bps} Beers, T. C. , Preston, G. W., \& Shectman, S. A. 1992, \aj, 103, 1987
\bibitem[]{bot01} Bode, P., Ostriker, J. P., \& Turok, N. 2001, \apj, 556, 93
\bibitem[b81]{b81} Bond, H. E. 1981, \apj, 248, 606
\bibitem[bcl99]{bcl99} Bromm, V., Coppi, P.S., \& Larson, R. B. 1999, 
\apj, 527, L5
\bibitem[bcl02]{bcl02} Bromm, V., Coppi, P.S., \& Larson, R. B. 2002, 
\apj, 564, 23
\bibitem[bfcl01]{bfcl01} Bromm, V., Ferrara, A., Coppi, P.S., \& Larson, R. B.
2001a, \mnras, 328, 969
\bibitem[bkl01]{bkl01} Bromm, V., Kudritzki, R. P., \& Loeb, A.
2001b, \apj, 552, 464
\bibitem[bl(2004)]{blar04} Bromm, V., \& Larson, R. B. 2004, \araa, preprint
(astro-ph/0311019)
\bibitem[bl(2002)]{bl02} Bromm, V., \& Loeb, A. 2002, \apj, 575, 111
\bibitem[bl(2003a)]{bl03a} Bromm, V., \& Loeb, A. 2003a, \apj, 596, 34
\bibitem[bl(2003b)]{bl03b} Bromm, V., \& Loeb, A. 2003b, \nat, 425, 812
\bibitem[byh]{byh} Bromm, V., Yoshida, N., \& Hernquist, L. 2003, \apj, 596, L135
\bibitem[]{c81} Carlberg, R. G. 1981, \mnras, 197, 1021
\bibitem[]{c87} Carr, B. J. 1987, \nat, 326, 829
\bibitem[]{cay03} Cayrel, R., et al. 2003, \aap, in press (astro-ph/0311082)
\bibitem[]{c03a} Cen, R. 2003a, \apj, 591, L5
\bibitem[]{c03b} Cen, R. 2003b, \apj, 591, 12
\bibitem[]{ch02} Christlieb, N., et al. 2002, \nat, 419, 904
\bibitem[cfw03]{cfw03} Ciardi, B., Ferrara, A., \& White, S. D. M. 2003, \mnras, 344, L7
\bibitem[cb03]{cb03} Clarke, C. J., \& Bromm, V. 2003, \mnras, 343, 1224
\bibitem[]{cr86} Couchman, H. M. P.,\& Rees, M. J. 1986, \mnras, 221, 53
\bibitem[]{DiM03} Di Matteo, T., Croft, R. A. C.,
Springel, V., \& Hernquist, L. 2003, \apj, 593, 56
\bibitem[]{db96} Draine, B. T.,\& Bertoldi, F. 1996, \apj, 468, 269
\bibitem[fan03]{fan03} Fan, X., et al. 2003, \aj, 125, 1649
\bibitem[]{fp94} Flores, R. A., \& Primack, J. R. 1994, \apj, 427, L1
\bibitem[]{fbh02} Freeman, K., \& Bland-Hawthorn, J. 2002, \araa, 40, 487
\bibitem[]{fc00} Fuller, T. M., \& Couchman, H. M. P. 2000, \apj, 544, 6
\bibitem[]{gp98} Galli, D., \& Palla, F. 1998, \aap, 335, 403
\bibitem[]{go97} Gnedin, N. Y., \& Ostriker, J. P. 1997, \apj, 486, 581
\bibitem[]{hh03} Haiman, Z.,\& Holder, G. P. 2003, \apj, 595, 1
\bibitem[]{hl97} Haiman, Z.,\& Loeb, A. 1997, \apj, 483, 21
\bibitem[]{hl01} Haiman, Z.,\& Loeb, A. 2001, \apj, 552, 459
\bibitem[]{hrl97} Haiman, Z., Rees, M. J., \& Loeb, A. 1997, \apj, 476, 458
\bibitem[]{htl96} Haiman, Z., Thoul, A. A., \& Loeb, A. 1996, \apj, 464, 523
\bibitem[]{hw02} Heger, A.,\& Woosley, S. E. 2002, \apj, 567, 532
\bibitem[]{hf01} Hernandez, X., \& Ferrara, A. 2001, \mnras, 324, 484
\bibitem[]{hf02} Hirashita, H., \& Ferrara, A. 2002, \mnras, 337, 921
\bibitem[]{hol03} Holder, G. P., Haiman, Z., Kaplinghat, M., \& Knox, L.
2003, \apj, 595, 13
\bibitem[]{hm89} Hollenbach, D., \& McKee, C. F. 1989, \apj, 342, 306
\bibitem[hu02]{hu02} Hu, E. M., et al. 2002, \apj, 568, L75
\bibitem[Hu(2002)]{hd02} Hu, W.,\& Dodelson, S. 2002, \araa,
    40, 171
\bibitem[]{hut76} Hutchins, J. B. 1976, \apj, 205, 103
\bibitem[]{jch01} Jang-Condell, H., \& Hernquist, L. 2001, \apj, 548, 68
\bibitem[]{kap03} Kaplinghat, M., Chu, M., Haiman, Z., Holder, G. P., Knox, L.,
\& Skordis, C. 2003, \apj, 583, 24
\bibitem[]{kr83} Kashlinsky, A., \& Rees, M. J. 1983, \mnras, 205, 955
\bibitem[]{kw02} Kitsionas, S., \& Whitworth, A. P. 2002, \mnras, 330, 129
\bibitem[kog03]{kog03} Kogut, A., et al. 2003, \apjs, 148, 161
\bibitem[]{LR00}Lamb, D. Q., \& Reichart, D. E. 2000, \apj, 536, 1
\bibitem[]{l69} Larson, R. B. 1969, \mnras, 145, 271
\bibitem[]{l98} Larson, R. B. 1998, \mnras, 301, 569
\bibitem[]{l02} Larson, R. B. 2002, \mnras, 332, 155
\bibitem[]{l03} Larson, R. B. 2003, Rep. Prog. Phys., 66, 1651
\bibitem[]{ls84} Lepp, S., \& Shull, J. M. 1984, \apj, 280, 465
\bibitem[]{lb01} Loeb, A., \& Barkana, R. 2001, \araa, 39, 19
\bibitem[]{lh97} Loeb, A., \& Haiman, Z. 1997, \apj, 490, 571
\bibitem[]{lr94} Loeb, A., \& Rasio, F. A. 1994, \apj, 432, 52
\bibitem[]{mba01} Machacek, M. E., Bryan, G., \& Abel, T. 2001, ApJ, 548, 509
\bibitem[m03]{mbh03} Mackey, J., Bromm, V., \& Hernquist, L. 2003, ApJ, 586, 1
\bibitem[ms86]{ms86} Mac Low, M.-M., \& Shull, J. M. 1986, \apj, 302, 585
\bibitem[]{mfr01} Madau, P., Ferrara, A., \& Rees, M. J. 2001, ApJ, 555, 92
\bibitem[]{mcd61} McDowell, M. R. C. 1961, Observatory, 81, 240
\bibitem[mes03]{mes03} Miralda-Escud\'{e}, J. 2003, Science, 300, 1904
\bibitem[]{mo99} Moore, B., et al. 1999, ApJ, 524, L19
\bibitem[]{mfr01} Mori, M., Ferrara, A., \& Madau, P. 2002, ApJ, 571, 40
\bibitem[man98]{man98} 
Motte, F., Andr\'{e}, P., \& Neri, R. 1998, \aap, 336, 150
\bibitem[]{nu99} Nakamura, F., \& Umemura, M. 1999, ApJ, 515, 239
\bibitem[]{nu01} Nakamura, F., \& Umemura, M. 2001, ApJ, 548, 19
\bibitem[]{nu02} Nakamura, F., \& Umemura, M. 2002, ApJ, 569, 549
\bibitem[]{noz03} Nozawa, T., Kozasa, T., Umeda, H.,
Maeda, K., \& Nomoto, K. 2003, ApJ, in press (astro-ph/0307108)
\bibitem[]{oh02} Oh, S. P., \& Haiman, Z. 2002, \apj, 569, 558
\bibitem[]{oh01} Oh, S. P., Nollett, K. M., Madau, P., \& Wasserburg, G. J. 2001,
\apj, 562, L1
\bibitem[]{om00} Omukai, K. 2000, ApJ, 534, 809
\bibitem[]{oi02} Omukai, K., \& Inutsuka, S. 2002, MNRAS, 332, 59
\bibitem[]{on98} Omukai, K., \& Nishi, R. 1998, ApJ, 508, 141
\bibitem[]{op01} Omukai, K., \& Palla, F. 2001, ApJ, 561, L55
\bibitem[]{op03} Omukai, K., \& Palla, F. 2003, ApJ, 589, 677
\bibitem[]{oy03} Omukai, K., \& Yoshii, Y. 2003, ApJ, in press
(astro-ph/0308514)
\bibitem[]{og96} Ostriker, J. P.,\& Gnedin, N. Y. 1996, \apj, 472, L63
\bibitem[]{p83} Palla, F., Salpeter, E. E., \& Stahler, S. W. 1983, ApJ, 271, 632
\bibitem[]{pd68} Peebles, P. J. E., \& Dicke, R. H. 1968, \apj, 154, 891
\bibitem[]{pei03} Peiris, H. V., et al. 2003, \apjs, 148, 213
\bibitem[]{ps99} Puy, D., \& Signore, M. 1999, NewAR, 43, 223
\bibitem[]{qsw02} Qian, Y.-Z., 
Sargent, W. L. W., \& Wasserburg, G. J. 2002, \apj, 569, L61
\bibitem[]{qw02} Qian, Y.-Z., \& Wasserburg, G. J. 2002, \apj, 567, 515 
\bibitem[rees76]{r76} Rees, M. J. 1976, \mnras, 176, 483
\bibitem[rees84]{r84} Rees, M. J. 1984, \araa, 22, 471
\bibitem[ro77]{ro77} Rees, M. J., \& Ostriker, J. P. 1977, \mnras, 179, 541
\bibitem[ro04]{ro04} Ricotti, M., \& Ostriker, J. P. 2004, \mnras,
submitted (astro-ph/0310331)
\bibitem[]{rip02} 
Ripamonti, E.,  Haardt, F., Ferrara, A., \& Colpi, M. 2002, \mnras, 334, 401
\bibitem[]{sf03} Salvaterra, R., \& Ferrara, A. 2003, \mnras, 339, 973
\bibitem[]{sbk02} Santos, M. R., Bromm, V., \& Kamionkowski, M. 2002, \mnras, 336, 1082
\bibitem[]{sz67} Saslaw, W. C., \& Zipoy, D. 1967, \nat, 216, 976
\bibitem[]{sbk02} Scannapieco, E., Schneider, R., \& Ferrara, A. 2003, \apj, 589, 35
\bibitem[]{sch02} Schaerer, D. 2002, \aap, 382, 28
\bibitem[]{sch03} Schaerer, D. 2003, \aap, 397, 527
\bibitem[]{}Schneider, R., Ferrara, A., Natarajan, 
P., \& Omukai, K. 2002, \apj, 571, 30
\bibitem[]{}Schneider, R., Ferrara, A., Salvaterra, 
R., Omukai, K., \& Bromm, V. 2003a, \nat, 422, 869
\bibitem[]{}Schneider, R., Ferrara, A., \& Salvaterra, R. 
2003b, \mnras, submitted (astro-ph/0307087)
\bibitem[sp53]{sp53} Schwarzschild, M., \& Spitzer, L. 1953, Observatory, 73, 77
\bibitem[sss]{sss} Seager, S., Sasselov, D., \& Scott, D. 1977, \apjs, 128, 407
\bibitem[sk]{sk} Shapiro, P. R., \& Kang, H. 1987, \apj, 318, 32
\bibitem[]{ss02} Shibata, M., \& Shapiro, S. L. 2002, \apj, 572, L39
\bibitem[s77]{s77} Silk, J. 1977, ApJ, 211, 638
\bibitem[s83]{s83} Silk, J. 1983, MNRAS, 205, 705
\bibitem[sr98]{sr98} Silk, J., \& Rees, M. J. 1998, \aap, 331, L1
\bibitem[]{sok03a} Sokasian, A., Abel, T., Hernquist, L., \& Springel, V. 2003a, 
\mnras, 344, 607
\bibitem[]{sok03b} Sokasian, A., Yoshida, N., Abel, T., Hernquist, L., \& Springel, V. 2003b, 
\mnras, submitted
(astro-ph/0307451)
\bibitem[]{sbl03} Somerville, R. S., Bullock, J. S., \& Livio, M. 2003, \apj, 593, 616
\bibitem[dns03]{dns03} Spergel, D. N., et al. 2003, \apjs, 148, 175
\bibitem[]{tmck03} Tan, J. C., \& McKee, C. F. 2003, ApJ, in press
(astro-ph/0307414)
\bibitem[teg97]{teg97} 
Tegmark, M., Silk, J., Rees, M. J., Blanchard, A., Abel, T., \& Palla, F.
1997, \apj, 474, 1
\bibitem[]{tf01} Todini, P.,\& Ferrara, A. 2001, \mnras, 325, 726
\bibitem[TS]{ts00} Tumlinson, J.,\& Shull, J. M. 2000, \apj, 528, L65
\bibitem[TS]{tsv} Tumlinson, J.,Shull, J. M., \& Venkatesan, A. 2003, \apj, 584, 608
\bibitem[]{ueh96} Uehara, H., Susa, H., Nishi, R., Yamada, M.,
\& Nakamura, T. 1996, \apj, 473, L95
\bibitem[]{un02} Umeda, H., \& Nomoto, K. 2002, \apj, 565, 385
\bibitem[]{un03} Umeda, H., \& Nomoto, K. 2003, \nat, 422, 871
\bibitem[]{wv03} Wada, K., \& Venkatesan, A. 2003, \apj, 591, 38
\bibitem[]{wc87} Wolfire, M. G., \& Cassinelli, J. P. 1987, \apj, 319, 850
\bibitem[]{wl03a} Wyithe, J. S. B., \& Loeb, A. 2003a, \apj, 586, 693
\bibitem[]{wl03b} Wyithe, J. S. B., \& Loeb, A. 2003b, \apj, 588, L69
\bibitem[]{wl03c} Wyithe, J. S. B., \& Loeb, A. 2003c, \apj, 590, 691
\bibitem[]{yon72} Yoneyama, T. 1972, \pasj, 24, 87
\bibitem[yos03a]{yos03a} Yoshida, N., Abel, T., Hernquist, L., \& Sugiyama, N.
2003a, \apj, 592, 645
\bibitem[ybh]{ybh04} Yoshida, N., Bromm, V., \& Hernquist, L. 2004, \apj,
submitted (astro-ph/0310443)
\bibitem[yos03b]{yos03b} 
Yoshida, N., Sokasian, A., Hernquist, L., \& Springel, V.
2003b, \apj, 591, L1
\bibitem[yos03c]{yos03c} 
Yoshida, N., Sokasian, A., Hernquist, L., \& Springel, V.
2003c, \apj, in press (astro-ph/0305517)
\bibitem[ys79]{ys79} 
Yoshii, Y., \& Sabano, Y. 1979, \pasj, 31, 505
\bibitem[ys86]{ys79} 
Yoshii, Y., \& Saio, H. 1986, \apj, 301, 587

\end{thebibliography}
\end{document}